\documentstyle[12pt]{article}
\title{}
\author{}
\begin{document}

\baselineskip=8 mm plus 1mm minus 1 mm

 \begin{center}
{\bf Space positronium detection by radio measurements}

\vspace*{.5cm}

V. Burdyuzha${}^{1}$, Ph. Durouchoux${}^{2}$, V. Kauts${}^{1}$

\vspace*{.3cm}

${}^{1}$ Astro Space Center Lebedev Physical Institute\\
of Russian Academy of Sciences,\\
Profsoyuznaya 84/32, 117810 Moscow, Russia\\
 email: burdyuzh@dpc.asc.rssi.ru, kauts@dpc.asc.rssi.ru\\
${}^{2}$ CE-Saclay, DSM, DAPNIA, Service d'Astrophysique,\\
 91191 Gif-sur Yvette Cedex, France\\
email: durouchoux@discovery.saclay.sea.fr

\vspace*{1cm}

Abstract
\end{center}

\vspace*{.3cm}

The possibility of positronium detection at radio wavelengths is investigated
in detail.
For orthopositronium, the fine structure lines of the $n = 2$ level
(1.62 cm, 2.30 cm and
3.48 cm) may be probably detected by radiotelescopes of next generations with
the sensitivity around $1 \mu Jy$. The spin-flip line of the
ground state (0.147 cm) may
be observed from sources with a narrow annihilation line by modern telescopes with
the sensitivity  around $10 \mu Jy$. In the
fine structure and spin-flip lines a maser effect is produced.
The expected radio fluxes in the spin-flip line
can reach detected limits for certain
source geometries and optimal physical conditions inside the source.

\vspace*{.3cm}


\vspace*{1cm}

\begin{center}
Introduction
\end{center}

The development of X-ray and $\gamma$-ray astronomy over the last ten years
has shown that the $(e^{+}e^{-})$ annihilation process is exceptionally
important in modern astrophysics; the presence in space of
processes leading to creation of positrons has been unambiguously
confirmed.
However, a narrow 511 KeV line has only been detected in the
Galactic center region (Teegarden, 1994; Cheng et al., 1997; Purcell et al.,
1997; Teegarden et al., 1997 and Smith, Purcell, Leventhal, 1997). There are
some sources, e.g. 1E 1740.7-2942,
Nova Muscae, and the Crab, whose spectra display bright
transient emission features near 0.5 MeV (see the review of Smith et
al., 1996). As theoretical analysis has shown, positron annihilation
proceeds mainly via positronium states if the temperature of
the annihilation medium is low (Gould, 1989; Guessoum, Ramaty,
Lingenfelter, 1991; Burdyuzha and Kauts, 1995 [hereafter BK]).

The probability of positronium detection in the UV ($\lambda = 2431 \dot{A}$)
and IR ($\lambda = 1.31 \mu m, 3.75 \mu m$) regions has been considered in
detail in our previous articles (Burdyuzha, Kauts, and Yudin, 1992;
Burdyuzha, Kauts, and Wallyn, 1996). Here we consider the possibility of
observing the radio line triplet at 8625 MHz (3.48 cm), 13010 MHz
(2.30 cm), and 18496 MHz (1.62 cm), corresponding to the fine structure
of the orthopositronium $n = 2$ level and the spin-flip line at
203387 MHz (0.147 cm) of the hyperfine structure of the ground state.

Since the observed narrow annihilation line sources are located in the
Galactic Center region where the absorption at UV and IR
wavelengths is high, the study of positronium detection at radio
wavelengths is very important.

\begin{center}
Positronium atom
\end{center}

The simplest positonium atom consists of an electron $e^{-}$ and a positron
$e^{+}$, and can exist in two states -- parapositronium with
zero total spin of its electron and positron, and orthopositronium with
the total spin equal to unity (see Fig. 1). The main distinction between
a positronium
and a hydrogen atom lies in the fact that in states with zero orbital momentum
positronium is an unstable atomic
system with lifetimes in the orthostate $\tau_{ortho} \simeq 1.33 \times
10^{-7} \; n^{3} sec$ and in the parastate $\tau_{para} \simeq 1.25
\times 10^{-10} \; n^{3} sec$ (here $n$ is the principal quantum
number). Also shown in Fig. 1 are the fine and hyperfine structure of the lower
levels of positronium. As in the case of the hydrogen atom, the formation
of positronium in an excited state leads to an electromagnetic cascade to
lower levels and to appearence of characteristic lines ($L_{\alpha}, L_{\beta}...,
H_{\alpha}, H_{\beta}..., P_{\alpha}, P_{\beta}...$).

The lifetimes of lower states to
annihilation are very short, so for low densities
an equilibrium population of levels has no
time to form and populations are determined by processes of
recombination or charge exchange followed by a radiative cascade and
annihilation. In these conditions the appearence of maser radiation seems
natural.

Positronium atoms are produced after the delay of positrons. The probability of
positronium formation on one positron ($f$) depends essentially on
the parameters of the medium (its temperature, the presence of dust, and so on).
The value of
$f$ in different media has been estimated by many authors
(e.g. Bussard, Ramaty, and Drachman, 1979) and is $\sim 0.9 \div 0.95$.
For the case of an ionized hydrogen medium the rate of radiative capture
in the process
$e^{+} + e^{-} \rightarrow Ps + \gamma$ is higher than the rate of
direct annihilation $e^{+} + e^{-} \rightarrow 2 \gamma$ at
temperatures $< 10^{6} \; K$ (Gould, 1989). For the case of an
atomic or molecular
hydrogen medium an energetic positron, after entering the medium,
effectively loses its energy by ionization and excitation of atoms and
molecules of hydrogen and by Coulomb collisions. In this case annihilation
processes proceed
at low energies and positronium atoms will be formed at these
energies by charge exchange  reactions before annihilation.

For astrophysical applications it is important that the formation of
excited states
of positronium be suppressed (BK). Therefore transitions between
lower levels are more promising for observations.

\begin{center}
Radio lines of the orthopositronium n=2 fine structure level
\end{center}

The triplet of radio lines corresponding to $n = 2$ fine structure transitions
may appear for radiative transitions $2^{3} S_{1} \rightarrow 2^{3} P_{i} \;\;
(i = 0, 1, 2)$ (see Fig 1). The lifetime
of the $2^{3} S_{1}$ level is determined by three-photon annihilation
$\tau (2^{3} S_{1}) = 1.1 \mu s$ while the lifetimes of the $2^{3} P_{i}$
levels are
determined by the $L_{\alpha}$ transition $\tau (2^{3} P_{i}) = 3.2 ns$. It is
easy to see that a strong overpopulation of the
$2^{3} S_{1}$ level is created and that maser emission
takes place in this triplet.

The one-dimensional radiative transfer equation of in a line of this triplet is:
$$\frac{dI^{i}_{\nu}}{dx} = \frac{h \nu_{i}}{\Delta \nu_{i}} \left[
\dot{n}_{ps} (2^{3}S_{1})
\frac{A(2^{3}S_{1}\rightarrow 2^{3}P_{i})/4 \pi+B(2^{3}S_{1}\rightarrow
2^{3}P_{i})I^{i}_{\nu}/c} {A(2^{3}S_{1} \rightarrow 2^{3}P_{i})+
\sum \limits_{i} B(2^{3}S_{1} \rightarrow 2^{3}P_{i})
I^{i}_{\nu} \Delta \Omega_{i}/c+W^{3\gamma}_{annih.}} \right.$$
$$\left.- \dot{n}_{ps}(2^{3}P_{i})\frac{B(2^{3}P_{i}\rightarrow 2^{3}S_{1})
I^{i}_{\nu}/c}{B (2^{3}P_{i}\rightarrow 2^{3}S_{1})I^{i}_{\nu}\Delta
\Omega_{i}/c+A_{L_{\alpha}}} \right], i = 0, 1, 2 \eqno(1)$$
\\
where $I^{i}_{\nu}$ is the intensity in one of the lines of
the triplet; $\dot{n}_{ps}$ is the rate of positronium formation in
the respective states
per unit volume; $A (2^{3} S_{1} \rightarrow 2^{3} P_{i})$ are
the probabilities of spontaneous transitions $5.71 \times 10^{-6}
s^{-1}, 5.96 \times 10^{-6} s^{-1}$, and $2.89 \times 10^{-6}
s^{-1}$ in levels $2^{3}P_{o}, 2^{3}P_{1}$, and $2^{3}P_{2}$, respectively,
and $B_{i}$ are the Einstein coefficients. $\Delta \Omega_{i}$ is the
degree of background anisotropy; $\;\;W^{3 \gamma}_{annih.}$ is the probability
of three-photon
annihilation with the $2^{3} S_{1}$ level and $A_{L_{\alpha}}$ is the probability
of the $L_{\alpha}$ transition.

In this expression collisional processes are not included
since densities in the annihilation region are low.
For the case of ionized hydrogen this leads to the limit
$n {\scriptscriptstyle
  {}^{<}_{\sim}} 10^{9} \sqrt{\rm {T}_{eV}} \; cm^{-3}$, where T is the Maxwell
temperature of ionized hydrogen (Burdyuzha,
Kauts, and Yudin, 1992). For the case of atomic and molecular hydrogen our
approximation
will be valid over a wider range of densities.

Since for any physical conditions the inequality $A(2^{3}S_{1}
\rightarrow 2^{3}P_{i})/4 \pi \ll
B(2^{3} S_{1} \rightarrow 2^{3} P_{i})I^{i}_{\nu}/c$ holds,
we obtain:
$$\frac{dI^{i}_{\nu}}{dx} = \frac{h \nu_{i}}{\Delta \nu_{i}} \left[
\dot{n}_{ps} (2^{3}S_{1})
\frac{B(2^{3} S_{1} \rightarrow 2^{3}P_{i})
I^{i}_{\nu}/c} {\sum \limits_{i=0,1,2} B(2^{3}S_{1} \rightarrow 2^{3}P_{i})
I^{i}_{\nu} \Delta \Omega_{i}/c + W^{3 \gamma}_{annih.}} - \right.$$

$$\left.- \dot{n}_{ps} (2^{3}P_{i}) \frac{B(2^{3}P_{i} \rightarrow 2^{3}S_{1})
I^{i}_{\nu}/c} {B (2^{3}P_{i} \rightarrow 2^{3}S_{1}) I^{i}_{\nu} \Delta
\Omega_{i}/c + A_{L_{\alpha}}} \right], i = 0, 1, 2 \eqno(2)$$
\vspace{0.01cm}

For the conditions $B(2^{3}S_{1} \rightarrow 2^{3}P_{i})
I^{i}_{\nu} \Delta \Omega_{i}/c \approx W^{3 \gamma}_{annih.}$ a
transition from the unsaturated to the saturated regime takes place. The
respective brightness temperatures are
$T^{i, cr.}_{b} = (1.5, 1.0, and 1.3) \times 10^{11} (\frac{4 \pi}
{\Delta \Omega_{i}}) \; K$. In the subsequent discussion we consider three
possible cases.

1) $T_{b} < T^{i,cr}_{b}$ (that is, the maser regime is not saturated).
The probability of positronium production in the $2S, 2P$ states in different
media has been considered many times (Nahar, 1984; Hewitt et al.,
1990; Biswas et al., 1991). Since $A_{L_{\alpha}} \gg W^{3 \gamma}_{annih.}$
and the formation rates
$\dot{n}_{ps} (2^{3}S_{1})$ and $\dot{n}_{ps} (2^{3}P_{i})$ differ
insignificantly for positronium formation energies (BK),
so $\frac{\dot{n}_{ps} (2^{3}S_{1})}{ W^{3 \gamma}_{annih.}} \gg
\frac{\dot{n}_{ps} (2^{3}P_{i})} {A_{L_{\alpha}}}$ and
equation (2) takes the form:
$$\frac{d I^{i}_{\nu}}{dx} = \frac{h \nu_{i}}{\Delta \nu_{i}} \left [
\dot{n}_{ps} (2^{3}S_{1}) \frac{1}{W^{3 \gamma}_{annih.}}
B(2^{3}S_{1} \rightarrow 2^{3}P_{i}) I^{i}_{\nu}/c \right]$$

The corresponding depths $\tau_{i}$ are:
$$\tau_{i} = \frac{h \nu_{i}}{\Delta \nu_{i}} \dot{n}_{ps}
(2^{3}S_{1}) \frac{1}{W^{3 \gamma}_{annih.}} \frac{c^{2} A
(2^{3}S_{1} \rightarrow 2^{3}P_{i})}{8 \pi h \nu^{3}_{i}} \; l,
i = 0, 1, 2 \eqno(3)$$
\\
where $l$ is a characteristic dimension of the annihilation region.

Thus we see that $\tau_{i} \sim A_{i}/ \nu^{3}_{i} \sim (2 J_{i} + 1)$,
that is  $\tau_{0} : \tau_{1} : \tau_{2} = 1 : 3 : 5$.

As the simplest example, we consider a spherical
annihilation line source radiating isotropically which has
dimension $l$
at a distance $R$ from the observer. We take $\dot{n}_{ps} (2^{3}S_{1}) \simeq
\dot{n}_{ps} (1^{3}S_{1})/8$
(for the case of ionized hydrogen the asymptotic behaviour
of these values was investigated by Burdyuzha et al. (1992);
for atomic and molecular hydrogen the detail analysis and
the rationale for  choosing this approximation was presented by BK).
Then for the most
intense line $2^{3}S_{1} \rightarrow 2^{3}P_{2}$ ($\nu_{2}$ = 8625 MHz)
the optical depth $\tau_{2}$ is
$$\tau_{2} \approx 5 \times 10^{-5} \left( \frac{0.75 f}{2-1.5 f} \right)
\left( \frac{F^{511}}{10^{-4}ph/cm^{2} s} \right)
\left( \frac{R}{10 \; kpc} \right)^{2} \left( \frac{l}{1 \;a.u.}
\right)^{-2} \eqno(4)$$
\\
for $\Delta \nu/ \nu \sim 1/200$ (the width depending on the motion of
positronium atoms (Teegarden, 1994 and BK) and the intrinsic line width
are of the same order).The value of
$\left( \frac{0.75 f}{2-1.5 f} \right) \sim 1-1.2$ for $f \sim 0.9-0.95$.

2) If $T_{b} > T^{i, cr}_{b}$, the maser regime is saturated
and its main characteristic will be a spectral flux density. In this
case the ratios of the spectral flux densities depend on the values
of $B (2^{3} S_{1} \rightarrow 2^{3} P_{i})I^{i}_{\nu} \Delta
\Omega_{i}$. The total photon fluxes in the lines of the triplet are
determined only by the rate of positronium formation in the $2^{3} S_{1}$
state:
$$F_{\nu}^{i} =\frac{\dot{n}(2^{3}S_{1}) l^{3}}{24 R^{2}} \frac{h
\nu_{i}}{\Delta \nu_{i}} \frac{B_{i} I^{i}_{\nu} \Delta
\Omega_{i}}{\sum \limits_{i} B_{i} I^{i}_{\nu} \Delta \Omega_{i}} \left(
\frac{4\pi}{\Delta \Omega_{F^{i}_{\nu}}} \right) \eqno(5)$$
\\
where $\Delta \Omega_{F^{i}_{\nu}}$ is the anisotropy of the
radio emission from the annihilation region in the respective line. As an
example, for a flat spectrum and $\Delta \Omega_{i} \simeq constant$ we have:
$$F_{\nu} (2^{3}S_{1} \rightarrow 2^{3} P_{2}) \approx
0.001 \left( \frac{0.75 f}{2-1.5 f} \right) \left( \frac{F^{511}}{10^{-4}
ph/cm^{2} s} \right)
\left(\frac{4 \pi}{\Delta
\Omega_{F^{2}_{\nu}}} \right) \;\; mJy \eqno(6)$$
\vspace{0.01cm}

3) For very high temperatures $T_{b} > T^{i, high}_{b} = (1.6, 3.3,
and 7.4) \times 10^{13} (\frac{4 \pi}{\Delta \Omega_{i}}) \; K$
determined by conditions $B (2^{3} P_{i}
\rightarrow 2^{3} S_{1}) I^{i}_{\nu} \Delta \Omega_{i}/c \approx
A_{L_{\alpha}}$, induced transitions dominate. Therefore
the ratios of the spectral flux densities are $F^{0}_{\nu} : F^{1}_{\nu} :
F^{2}_{\nu} \simeq 1 : 3 : 5$. For $T_{b} \ge T^{i, high}_{b}$
line broadening due to induced transitions is the most important effect
and an following
increase in temperature results in overlapping lines.
The spectral flux densities in this regime will also be determined
by formula (6) but it is necessary to take the
increase in line widths into account adding the coefficient
$ \sim (T^{i,high}_{b}/T_{b})$ to this formula.

In parapositronium the situation is more complex since the lifetimes of the

levels $\tau (2^{1} P_{1}) \approx \tau (2^{1} S_{o})$ and accurate
predictions require further study.

\begin{center}
The positronium spin-flip radio line
\end{center}

The hyperfine structure transition $1^{3}S_{1} \rightarrow 1^{1}S_{0}$
of the ground state (spin-flip line) and the fine structure transitions
can be observed in emission since the lifetime of the level
$1^{3}S_{1} (\tau = 1.33 \times 10^{-7} \; sec)$ is much longer than the
lifetime of the $1^{1} S_{o}$ level $(\tau = 1.25 \times 10^{-10} \;
sec)$. Here the maser effect is also evident. This transition is of the
$M1$ type with $\nu = 203387 \; MHz, \; A = 3.37 \times 10^{-8} \;
s^{-1}$ and intrinsic line width $\Delta \nu = 1.3 \times 10^{9}
\; Hz$. We have repeated the analysis for the spin-flip line similar
to previous one. The transition from the unsaturated to the saturated regime
occurs for $B (1^{3} S_{1} \rightarrow 1^{1} S_{o})
I_{\nu} \Delta \Omega/c \approx W^{3 \gamma}_{annih.} (1^{3} S_{1})$.
The corresponding brigthness temperature is $T^{cr}_{b} = 2.2 \times
10^{15} (\frac{4 \pi}
{\Delta \Omega}) \; K$. The optical depth in the unsaturated regime for
the conditions described in formula (4) is
$$\tau \approx 5 \times 10^{-11} \left( \frac{0.75 f}{2-1.5 f}
\right) \left( \frac{F^{511}}{10^{-4} ph/cm^{2} s}
\right)
\left( \frac{R}{10 \; kpc} \right)^{2} \left( \frac{l}{1 \;
a.u.} \right)^{-2} \eqno(7)$$
\vspace{0.01cm}

The spectral flux density in saturated regime from an annihilation
region is
$$F_{\nu} \approx 0.013 \left( \frac{0.75 f}{2-1.5 f} \right) \left(
\frac{F^{511}}{10^{-4} ph/cm^{2} s} \right)
\left( \frac{4 \pi}{\Delta \Omega_{F_{\nu}}} \right)
\; mJy \eqno(8)$$
\vspace{0.01cm}

The width depending on the motion of Ps atoms (Teegarden, 1994; BK)
and the intrinsic line width are of the same order
just as in the case of the fine structure lines.

If $T_{b} \ge T_{b}^{high} = 7.7 \times 10^{17} (\frac{4
\pi}{\Delta \Omega}) \; K$, which is determined by the condition $B (1^{1}
S_{o} \rightarrow 1^{3} S_{1}) I_{\nu} \Delta \Omega/c \approx W^{2
\gamma}_{annih.} (1^{1} S_{o})$, then the spectral flux density in
this regime will also be given by formula (8) but due to the
increase of the line width the coefficient
$(3T^{high}_{b}/4T_{b})$ appears.

\begin{center}
Conclusions
\end{center}

Simultaneous observations of triplet radio lines can solve the problem
of positronium identification. It follows from equation 4 that observations
of such lines in the unsaturated regime are possible only from
sufficiently compact and distant sources with the large flux in
an annihilation line. Today such sources are unknown. From equation 6,
the spectral flux densities in the saturated regime are just detectable with modern
radio telescopes for $\left( \frac{F^{511}}{10^{-4}
ph/cm^{2} s} \right) \left(\frac{4 \pi}{\Delta
\Omega_{F^{2}_{\nu}}} \right) \ge 1$. Note that formula 6 gives the total
spectral flux density from the entire annihilation region, and
that the flux may be larger if the source is anisotropic.  But it is certainly
an unnatural proposal.

For the spin-flip line the optical depth in the unsaturated regime is
very small since this is the M1 type transition and it is
impossible to detect. But the situation is different in the saturated regime.
The spectral flux density in the saturated regime is higher than that in
the triplet lines (equation 8) but the point of transition
from the unsaturated to the saturated regime ($T^{cr}_{b}$) is shifted.
As was shown by Kardashev (1979), this transition is optimal for
interstellar communication but so far attempts to observe possible
planetary systems at 203387 MHz have not been successful
(Mauersberger et al., 1995). It is very desirable to extend these
observations to possible sources of the 511 keV annihilation line with high
brightness temperatures in the mm range (radio jets sources, radiopulsars
and so on).
The search for this radio line can be done in the direction of the Galactic
center but to date a narrow annihilation line point source has not been
detected there, so the exact predictions cannot be done.
One promising source for spin-flip detection is radiopulsars.
The problem is a difficult one, since necessary conditions are
the presence of positrons (e.g., in a pulsar
wind), and an appropriate deceleration length;
a medium with a define density and dimension is required.

In addition to radio observations positronium can be also  detected
in the recombination
lines $L_{\alpha} (2431 \dot{A})\;$, $H_{\alpha} (1.31 \mu m)$ and
$P_{\alpha} (3.75 \mu m)$. This possibility was investigated in our
previous articles (Burdyuzha, Kauts, and Yudin, 1992; BK;
Burdyuzha, Kauts, and Wallyn, 1996). The $L_{\alpha}$ line is difficult to
observe in the direction of the Galactic center because of high absorption
but it may be observed from active galactic nuclei with UV bumps in
their spectra and from new point sources in our Galaxy which have
to be detected. Observations of $L_{\alpha} (2431 \dot{A})$ can be carried
out with the HST. The IR lines $H_{\alpha} (1.31 \mu m), \; H_{\beta} (0.97
\mu m)$ and $P_{\alpha} (3.75 \mu m)$ can be observed by Keck telescope
and probably by UKIRT, IRTF IR telescopes.

\begin{center}
Acknowledgements
\end{center}

We are grateful to V.Dubrovich, Ya.Istomin, N.Kardashev and V.Slysh for
useful discussions.

\vspace*{.5cm}

\begin{center}
Bibliography
\end{center}

\vspace*{.3cm}
\noindent
1. Biswas P. K., Mukherjee T., and Ghosh A. S., 1991, J. Phys. B. 24, 2601\\
2. Burdyuzha V. V., Kauts V. L., and Yudin N. P., 1992, A\&A 255, 459\\
3. Burdyuzha V. V., and Kauts V. L., 1995, Astronomy Letters 21, 159 \\
4. Burdyuzha V. V., Kauts V. L., and Wallyn P., 1996, A\&AS 120, C365\\
5. Bussard R. W., Ramaty R., and Drachman R. J., 1979, ApJ 228, 928\\
6. Cheng L.X., Leventhal M., Smith D.M., Purcell W.R. et al. 1997, ApJ 481, L43\\
7. Gould R. F., 1989, ApJ 344, 239\\
8. Guessoum N., Ramaty R., and Lingenfelter R. E., 1991, ApJ 378, 170\\
9. McClintock J. E., 1984, ApJ 282, 291\\
10.Nahar S., 1984, Phys. Rev. A 40, 6231\\
11.Hewitt R. N., Noble C. J., and Dransden D. H., 1990, J.Phys.B.At.Mol.Opt.
Phys 23, 4185\\
12.Kardashev N. S., 1979, Nat 278, 28 \\
13.Mauesberger R., Wilson T. L., Rood R. T., Bania T. M. et al., 1995,
A\&A 306, 141   \\
14.Purcell W. R., Dixon D. D., Cheng L. X., Leventhal M. et al., 1997,
Proceedings 22nd INTEGRAL Workshop "The Transparent Universe", St.Malo,
France, 16-20 Sept. 1996, ESA SP-382, 67\\
15.Smith D. M., Leventhal M., Cavallo R., Gehrels N., Tueller J., and
Fishman G., 1996, ApJ 471, 783\\
16.Smith D.M., Purcell W.R., Leventhal M., 1997, Proceedings of the fourt
Compton Symposium, Williamsburg, USA. Eds: Dermer D., Strickman M.,
Kurfess J. , 208 \\
17.Teegarden B. J.,1994, ApJS 92, 363\\
18.Teegarden B.J., Cline T.L., Gehrels N., Ramaty R. et al., 1997,
Proceedings of the fourth Compton Symposium, Williamsburg, USA. Eds:
Dermer D., Strickman M., Kurfess J., 1007\\

\end{document}